\documentclass[aps,twocolumn,superscriptaddress,pre]{revtex4-1} 

\usepackage{scrextend}
\usepackage{amssymb}  
\usepackage{amsmath}   
\usepackage{natbib}
\usepackage{epsfig}   
\usepackage{soul}
\usepackage{relsize}
\usepackage{graphicx,subfigure}   
\usepackage{dcolumn}
\usepackage{bm}  
\usepackage{soul}
\usepackage[english]{babel}
\usepackage[latin1]{inputenc}
\usepackage{xcolor}
\usepackage{comment}
\usepackage[update]{epstopdf}
\emergencystretch=\maxdimen
\begin{document}

\title{ Heat flux direction controlled by power-law oscillators \\under non-Gaussian fluctuations
}

\author{E. H.  Colombo} \email{ecolombo@ifisc.uib-csic.es}
\affiliation{IFISC (CSIC-UIB), Campus Universitat Illes Balears, 07122, Palma de Mallorca, Spain}

\author{L. A. C. A. Defaveri} \email{lucianno-defaveri@puc-rio.br}
\affiliation{Department of Physics, PUC-Rio, Rio de Janeiro, Brazil}

\author{C. Anteneodo} \email{celia.fis@puc-rio.br}
\affiliation{Department of Physics, PUC-Rio, Rio de Janeiro, Brazil}
\affiliation{Institute of Science and Technology for Complex Systems, Rio de Janeiro, Brazil}

\begin{abstract}
Chains of particles coupled through anharmonic interactions 
and subject to non-Gaussian baths can exhibit paradoxical outcomes 
such as heat currents flowing from colder to hotter reservoirs. 
Aiming to explore the role of generic non-harmonicities in mediating the contributions 
of non-Gaussian fluctuations to the direction of heat propagation, 
we consider a chain of power-law oscillators, with interaction potential 
$V(x) \propto |x|^\alpha$, subject to Gaussian and Poissonian baths at its ends.  
Performing numerical simulations and addressing heuristic considerations, 
we show that a deformable potential has bidirectional control over heat flux.
\end{abstract}

\maketitle

\section{Introduction}
Traditional concepts of equilibrium thermodynamics face important challenges in the 
realm of nonequilibrium processes. 
For instance,  theoretical and experimental studies of open systems 
have lead to review the definitions of fundamental quantities such as heat and work, circumventing apparent violations of standard laws~\cite{Parrondo2015,Niedenzu2018,Li2003,landi2013,kanazawa2012}. 

In particular, the exploration of heat conduction through a medium 
connected to baths with generalized properties, beyond Gaussian fluctuations, 
has revealed counter-intuitive phenomena, 
which called for the investigation on how concepts such as heat flux  
and temperature itself should be read in this 
scenario~\cite{kanazawa2012,kanazawa2013,morgado14,morgado16}. 
Assuming that the source of heat is non-Gaussian 
(thus described by an infinite set of cumulants as stated by the Marcinkiewicz  theorem~\cite{risken}), 
one finds that each cumulant can be interpreted as a source of stochasticity 
that plays a role in heat flux~\cite{kanazawa2012}. 
Importantly, the role of higher-order cumulants (beyond the second-order one) 
is not extrinsic but mediated by system properties~\cite{kanazawa2013,candido17}. 

\begin{figure}
\includegraphics[width=\columnwidth]{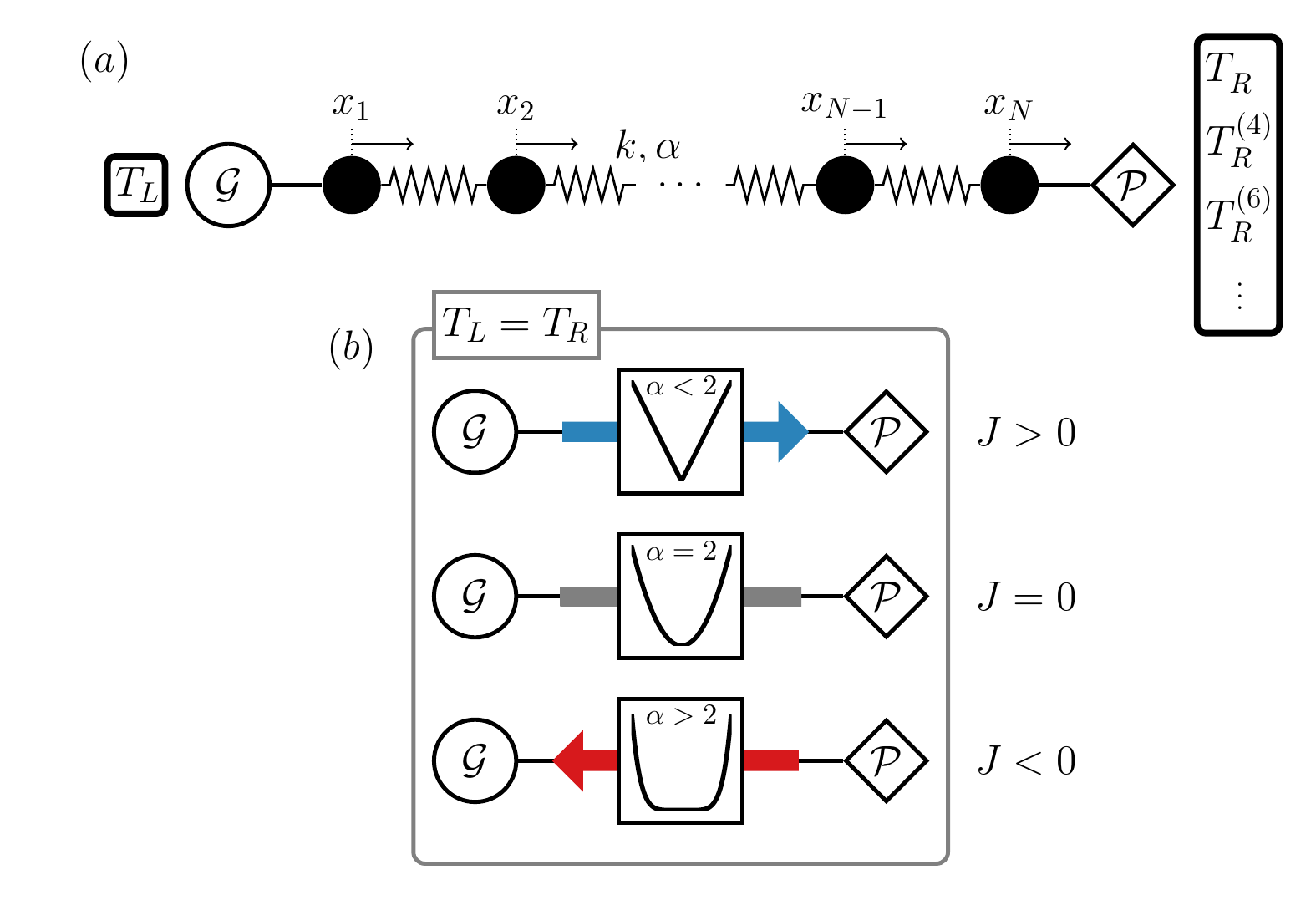}
\caption{(a) Chain of oscillators coupled at its ends to Gaussian and  Poissonian 
baths  at standard temperatures $T_L$ and $T_R$, respectively, but in the Poissonian 
case there are additional sources of stochasticity. 
Parameters $k$ and $\alpha$ control the intensity and nonlinearity of the interactions, 
respectively, the variable  $x_i$ measures the displacement of the $i$th particle 
from its equilibrium position. (b) Visual abstract of the main results on heat transport, representing the consequences of subharmonic ($\alpha<2$), harmonic ($\alpha=2$), superhamonic ($\alpha>2$) potentials to heat flux direction in the absence of temperature gradient.
}
\label{fig:chain}
\end{figure}

For a one-dimensional chain, within the classical framework of Fourier's law~\cite{fourier1822}, 
heat flux direction is an extrinsic property, established by the thermal baths.
Namely, the flux $J$ is given by
\begin{equation} \label{fourier}
J = \kappa \Delta T =\kappa (T_{L} - T_{R}) \, ,
\end{equation}
where $\Delta T$ is the temperature difference between the left and right baths 
(see Fig.~\ref{fig:chain}) 
and only the conductance $\kappa$ is characteristic of the propagation medium. 
The validity of the Fourier's law in one-dimensional systems 
has been a constant matter of 
debate~\cite{lieb1967,casher1971,lepri1997,Li2003,landi2013,lepri2003,
Das2014,Hurtado2016,olivares16}, concerning anomalous temperature profiles and divergence of 
the conductivity in the thermodynamic limit, even in equilibrium Gaussian scenarios. 
When plugging athermal (non-equilibrium) baths, exhibiting non-Gaussian statistics, 
still more drastic apparent violations of Fourier's law can emerge.
As a matter of fact, non-Gaussianity allied to nonlinear coupling of the chain elements, was shown to 
break down the classical picture for the direction of heat flow 
producing counter-intuitive 
heat transfer from colder to hotter reservoirs~\cite{kanazawa2013,candido17}. 
A granular motor, as previously studied in Refs.~\cite{gnoli2013,eshuis2010}, is a concrete example of a system that exhibits non-Gaussian features, being a suitable candidate for a nonequilibrium bath~\cite{kanazawa2015,sano2016}.
For this case, the collisions between bath particles and system contact can be modeled as a Poisson shot noise.

In a broad view of the problem, this coupling produces an effective temperature difference 
defined by a complex interplay between the system and the surrounding baths.
It has been previously shown, that in the particular case of a chain of particles 
coupled via the Fermi-Pasta-Ulam-Tsingou (FPUT) potential  
$V(x)= \frac{1}{2} k_1x^2 + \frac{1}{4}k_3 x^4$, with $k_1 \gg k_3>0$,   
 subjected to baths at each end obeying Poisson shot noise and Gaussian statistics, 
respectively,  there is a correction to the flux through the chain with respect to 
the Gaussian-Gaussian case~\cite{candido17,kanazawa2013}. 
This correction is unidirectional, 
from the Poisson to the Gaussian bath (see Fig.~\ref{fig:chain}), 
reflecting the interplay between nonlinearity and non-Gaussianity.
Nevertheless, a systematic investigation of the general role of nonlinearity 
in mediating the contributions of non-Gaussian fluctuations in heat transport is still lacking.

Then, we consider an interaction potential of the power-law form 

\begin{equation}
V_\alpha (x) =   k\frac{|x|^{\alpha}}{\alpha}\, ,
\label{eq:potential}
\end{equation}
where $k\ge 0$ and $\alpha$ is a real parameter, yielding the force  
\begin{equation}
F_\alpha(x) =  -k|x|^{\alpha-2}x. 
\label{eq:force}
\end{equation}
We consider $\alpha \in [1,+\infty)$, 
that includes two classes of nonlinearities:   
for $\alpha>2$ ($\alpha<2$) the force is super(sub)-linear with the displacement $x$, 
while the harmonic interaction is given by $\alpha=2$.

Equation~(\ref{eq:potential}) allows to scan between the paradigmatic cases of 
triangular potential, whose periodic extension 
(saw-tooth) is used in ratchet modeling ($\alpha=1$), 
and infinite square well ($\alpha \to \infty$); including the harmonic and quartic anharmonicities.  
Generic values of $\alpha$, not necessarily integer, 
can mimic realistic scenarios beyond 
simple harmonic oscillations, emerging due to nonlinear responses,  
at the macroscopic or atomic level~\cite{dnacurve,schmelcher2018,nlmachine,dhar2019}.  

The relevant features that we will discuss occur around $\alpha=2$, 
for which the stiffness, $k_\alpha(x) = k |x|^{\alpha-2}$ 
(such that $F_\alpha(x)=-k_\alpha(x) x$)  
goes to zero (${\alpha>2}$) or diverges (${\alpha<2}$) at the origin. 
We restrict our study to the region ${\alpha\ge 1}$ in order to avoid 
a divergent force at the origin. 
Actually,  if there were interest in investigating the region $\alpha<1$, like 
in the case $\alpha \to 0$ (logarithmic potential), 
studied in the context of optical lattices~\cite{barkai2010}, 
the potential given by Eq.~(\ref{eq:potential}) can be regularized  [see Eq.~(\ref{regular})]. 

Our results reveal that the role of higher-order cumulants 
is critically determined by the potential shape, which promotes negative 
corrections (for $\alpha>2$), as reported for the case of FPUT chains~\cite{kanazawa2013}, 
or positive ones (for $\alpha<2$). Thus,  a deformable potential can 
fully control the flux direction. By investigating the heat flux statistics, 
we show in detail that the effect arises 
from the competition between frequent (small) and rare (large) flux fluctuations, 
which is ruled by $\alpha$. 

We begin by defining the system, in the following section (Sec.~\ref{sec:model}). 
Next, in Sec.~\ref{sec:results}, we present the numerical results for the statistics of heat 
flux, which are accompanied by analytical considerations discussed in Sec.~\ref{sec:heuristic}. 
Lastly, in Sec.~\ref{sec:final}, we address final remarks.

\section{System}
\label{sec:model}
 
We consider a one-dimensional chain of nonlinear oscillators coupled 
to Gaussian and Poisson baths at each extremity. 
A pictorial representation of the system is given in Fig.~\ref{fig:chain}a.

The Hamiltonian of the chain is 
\begin{equation}\label{hamiltonian}
H =  \sum_{i=1}^N \left[ \frac{1}{2m} p_i^2 + \sum_{i=1}^{N-1} V_\alpha(x_{i+1}-x_i) \right] \,,
\end{equation} 
where $x_i$ and $p_i$ represent the displacement and momentum of the $i$-th particle in the lattice, 
respectively, and the potential $V_\alpha$.    
Therefore, the equations of motion for the central particles $(2,\ldots,i,\ldots,N-1)$ are 

\begin{equation}\label{model_central}
m \, \ddot{x}_i  =  F_\alpha(x_i - x_{i+1}) + F_\alpha(x_i - x_{i-1})    \,,
\end{equation}
where $m$ is the mass of the oscillators. The equations for the particles in contact with the baths  read   

\begin{eqnarray}\label{model_ends}
m \, \ddot{x}_1 + \gamma \dot{x}_1 &=& F_\alpha (x_1 - x_2) + \xi_L (t) \,, \\ 
m \, \ddot{x}_N + \gamma \dot{x}_N &=& F_\alpha (x_N - x_{N-1}) + \xi_R (t)\,, 
\end{eqnarray}
where $\gamma$ is the friction coefficient, and $\xi_L$ and $\xi_R$ are noises 
that mimic the injection of stochasticity by the baths.

On the left end of the chain, we connect a standard delta-correlated thermal bath $\mathcal{G}$ 
with Gaussian statistics.
Therefore, its cumulants are given by
$\langle \xi_L(t_1)\ldots \xi_L(t_n)\rangle_c = K_n^{(L)}\prod_{i=2}^{n}\delta(t_i-t_{i-1})$, where 
\begin{align}
\label{gauseq}
K_n^{(L)} =
\begin{cases} 
      2\gamma T_L, & n=2\, , \\
      0,           & n\neq 2\, .
   \end{cases}  
\end{align}

On the right end, a generalized bath, with an infinite number of cumulants, is introduced. 
It is implemented by a Poissonian symmetric shot noise $\mathcal{P}$, 
setup by a series of delta-correlated instantaneous (negative and positive) 
force pulses with exponentially distributed amplitude $\Phi_i$ and time lag $\tau_i=t_{i+1}-t_i$, such that
\begin{equation}
\xi_R(t) = \sum_i \Phi_i \delta(t-t_i)\, ,
\end{equation}
with
\begin{equation}
p_t(\tau_i) = \lambda e^{-\lambda \tau_i}\quad\text{and}
\quad p_\Phi(\Phi_i)= \bar{\Phi}^{-1}e^{-|\Phi_i|/\bar\Phi} \,,
\end{equation}
where $\lambda$ is the shot rate and $\bar{\Phi}$ is the average absolute value of the amplitude.
Under this definition, the cumulants are 
$\langle \xi_R(t_1)\ldots \xi_R(t_n)\rangle_c = K_n^{(R)}\prod_{i=2}^{n}\delta(t_i-t_{i-1})$, where
\begin{align}
\label{poiseq}
K_n^{(R)} =
\begin{cases} 
      \lambda n! \bar\Phi^n,   & \mbox{even $n\geq 2$} \,, \\		
					          0,         & \mbox{otherwise} \,.   
   \end{cases}
\end{align}

The discrete character of the bath generates an infinite set of cumulants that constitute, as previously demonstrated~\cite{morgado12,kanazawa2012,kanazawa2013}, 
sources of stochasticity that have been associated 
to the concept of higher-order temperatures, 
namely $T_R^{(n)} \propto  K_n^{(R)}=\lambda n! \bar\Phi^n$~\cite{morgado16}.  
In particular, the canonical temperature is given by 
$T_R \equiv T_R^{(2)}= K_2^{(R)}/(2\gamma)=
\lambda\bar\Phi^2/\gamma$.
This constitutes a fluctuation-dissipation relation, analogous to Eq.~(\ref{gauseq}).

\section{Numerical results}
\label{sec:results}
 
We integrate the equations of motion 
by means of a  stochastic Runge-Kutta scheme~\cite{RK4}, 
setting $m=\gamma=k=1$.
As initial conditions, particles are at their equilibrium positions with zero velocities, 
that is, $x_i=p_i=0$.
We use $\bar\Phi=1/2$ and $\lambda=  \gamma T_R/\bar{\Phi}^2$ for the Poisson bath. 
We restrict our present study to the case where the standard temperatures at the ends 
coincide ($T_R=T_L$),  
in order to highlight the phenomena that emerge when nonharmonicity and generalized bath properties are put together.


\begin{figure}[b]
\includegraphics[width=\columnwidth]{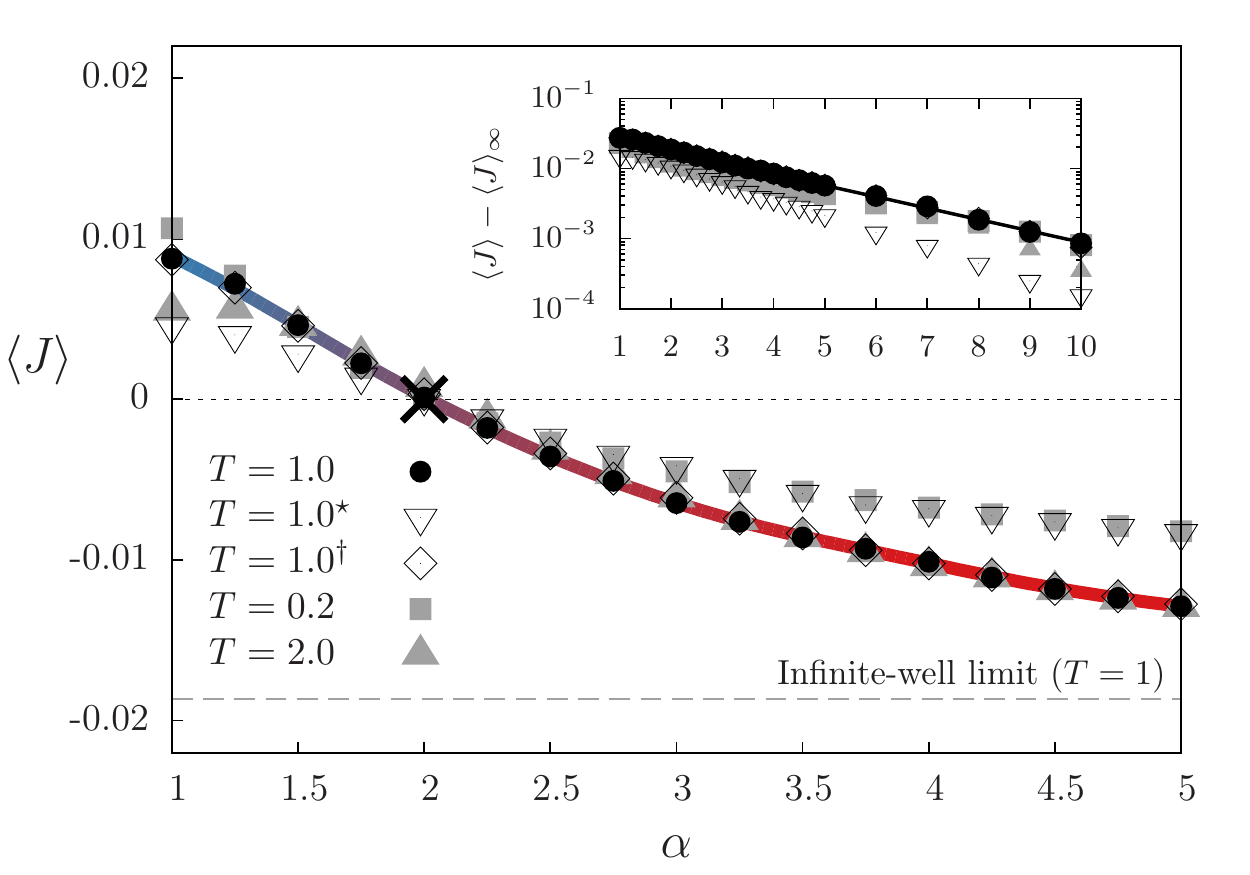}
\caption{Average flux $\langle J \rangle$, 
for a three-particle chain with $\Delta T = 0$ and $T=T_L=T_R$ (indicated in the figure), 
considering   Gaussian (left) and Poissonian (right) baths, 
for different values of $\alpha$ in Eq.~(\ref{eq:potential}). 
The ``$\times$'' symbol represents the harmonic case for which $\langle  J \rangle =0$. 
The solid colored line highlights flux direction for the $T=1$ case, 
according to Fig.~\ref{fig:chain}.
Horizontal lines highlight the zero flux level (dotted) 
and the infinite-well limit (dashed) for $T=1$. 
Results for the regularized potential in Eq.~(\ref{regular}),
with  $x_0=0.1^\dagger$, $1.0^\star$, are also shown for comparison. 
The inset shows the excess flux with respect to the infinite-well limit 
(obtained numerically by a scaling procedure),  
putting into evidence an exponential decay (solid black line) towards the limit level. 
}
\label{fig:gpflux}
\end{figure}

Heat transport is analyzed by means of the average flux that passes through the chain. 
In the long-time regime, when a steady state is attained, 
the injected and rejected heats are the same. 
Thus, the flux that leaves the system can be written as  
\begin{equation}
\label{instJ}
J = F_\alpha(x_N-x_{N-1})(v_N+v_{N-1})/2 \, ,
\end{equation}
and its time averaged value is  
\begin{equation}
\langle J \rangle = \lim_{t\to\infty}\frac{1}{t} \int_0^t J(t) \,\text{d}t\, .
\end{equation}

Figure~\ref{fig:gpflux} displays the behavior of the mean current as a function of 
$\alpha$ exponent, when  $\Delta T =0$. 
For the harmonic confinement ($\alpha=2$),   
there is no flux, while transport occurs otherwise, as a result 
of the intertwining between system and bath properties.
For $\alpha>2$, the flux becomes negative (i.e., heat flows from right to left), 
as if the effective temperature of the Poisson bath were larger than $T_R$
given by its second cumulant, due to the effect of the higher-order ones. 
Contrastingly, for $\alpha<2$, the role of this set of higher-order cumulants is reverted, 
reducing the effective temperature, and hence the flux direction. 
Thus, a deformable potential (around $\alpha=2$) has bidirectional control over heat transfer.

\begin{figure}[h]
\includegraphics[width=\columnwidth]{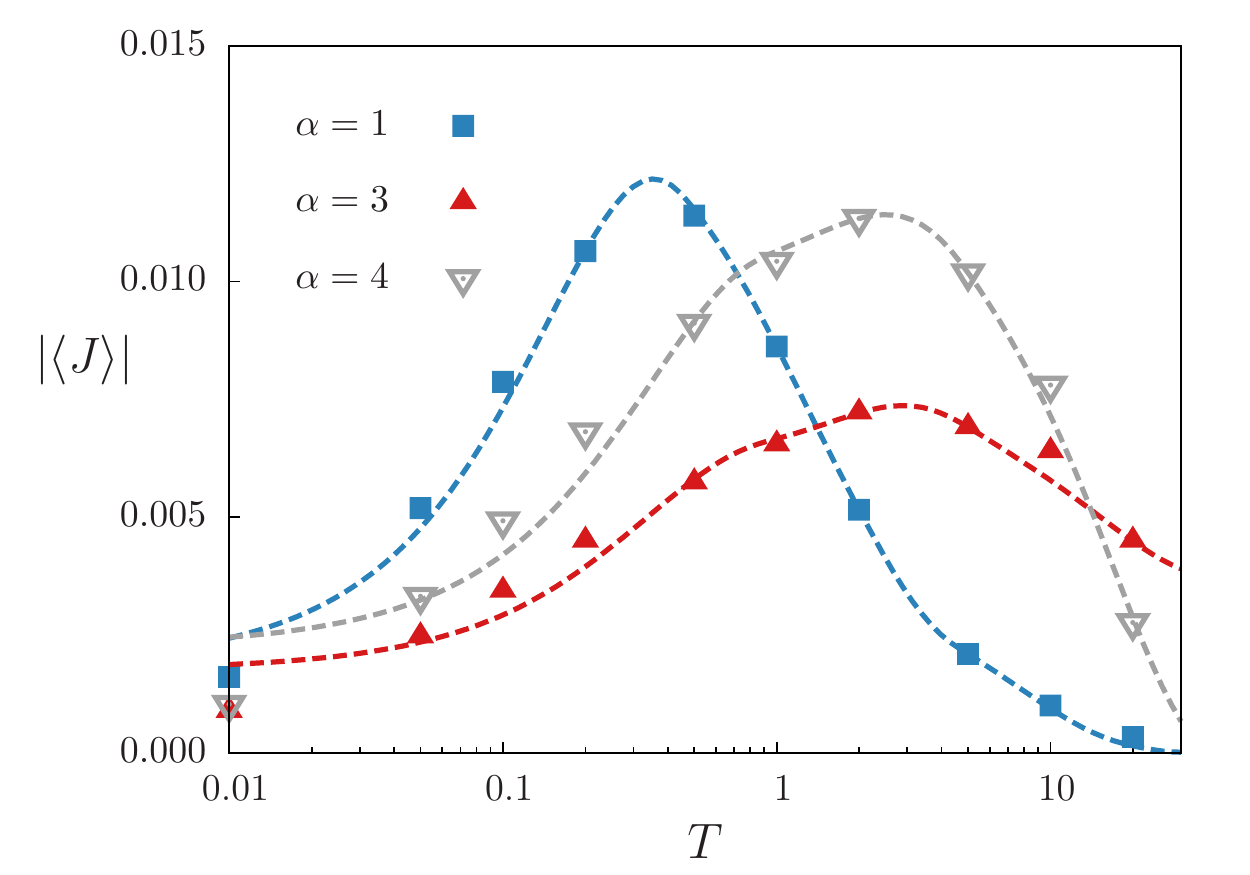}
\caption{Average flux magnitude $|\langle J \rangle|$, 
for a three-particle chain with $\Delta T = 0$ as a function of $T=T_L=T_R$ 
[varying $\lambda$, see Eq.~(\ref{poiseq})], 
considering  Gaussian (left) and Poissonian (right) baths, 
for $\alpha$ corresponding to the sub- and super-harmonic cases,  
in  Eq.~(\ref{eq:potential}). 
The dashed lines are guides to the eye.}
\label{fig:temp}
\end{figure}

In Fig.~\ref{fig:gpflux}, we show in detail that this nonlinear control is robust against the temperature level $T$ ($=T_L=T_R$) to which the chain is subjected, but, its impact is maximized for an intermediate value of $T$, as depicted in Fig.~\ref{fig:temp}.

These results put into evidence the impact of the type of confinement in heat flux direction. 
However, it is not yet clear which are the regulatory mechanisms for the emergence of this interplay. 
We will show that, in the present scenario, flux direction is controlled by a competition between rare 
(large scale) and frequent (short scale) heat transport events, controlled 
by $\alpha$.

\begin{figure}[t]
\includegraphics[width=0.95\columnwidth]{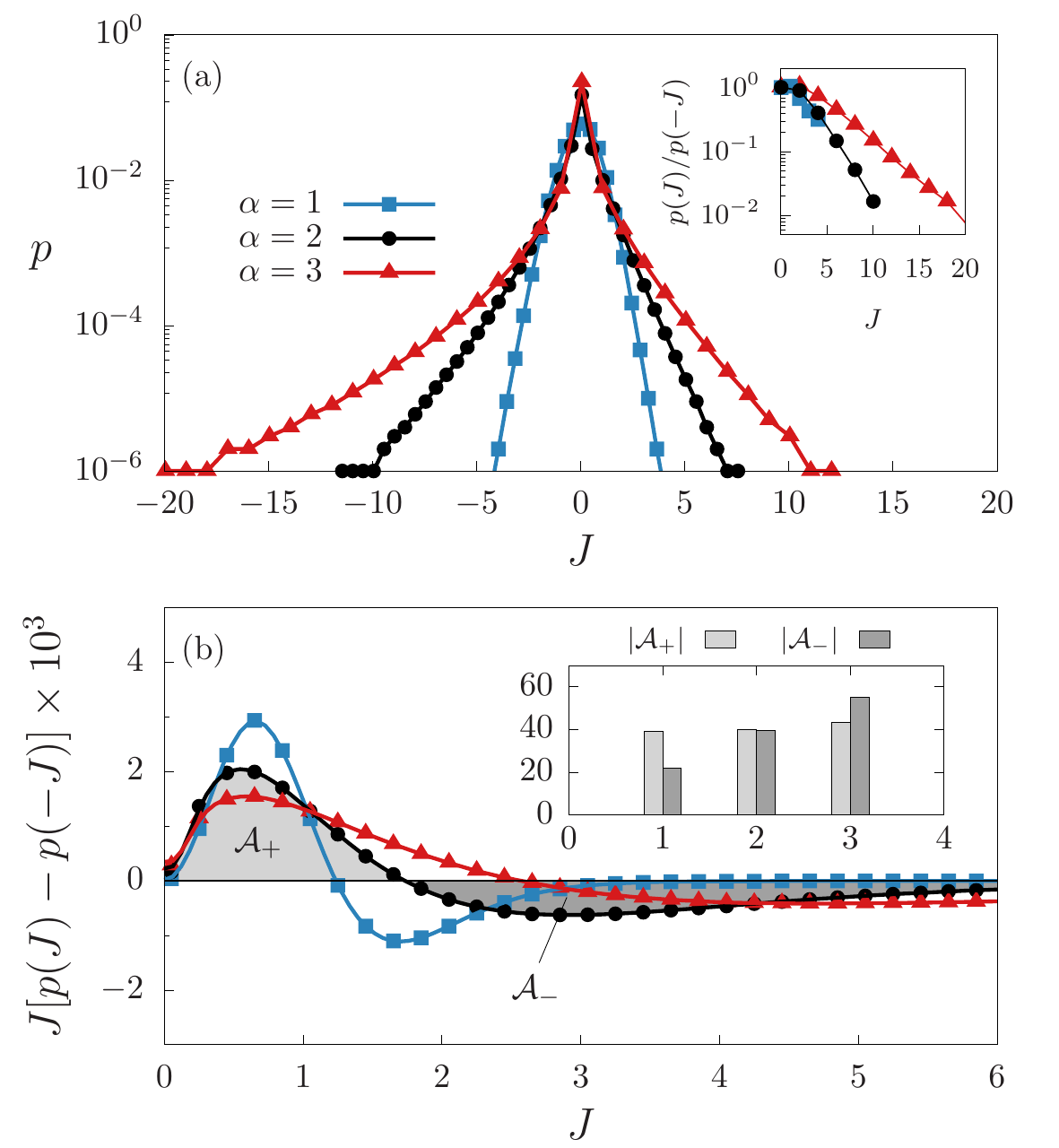}\\
\caption{(a) Probability density function $p(J)$ for the instantaneous flux 
$J=F_\alpha(x_{i+1}-x_i)(v_i+v_{i+1})/2$, 
for a three particle chain with  $T_L=T_R=1$, subject to Gaussian-Poisson baths, 
for different values of $\alpha$. 
Inset: asymmetry ratio vs. $J$.  
(b) Contribution to the average flux as a function of positive $J$, for the same parameters. 
Inset: absolute value of the short scale ($\mathcal{A}_+$) and large scale ($\mathcal{A}_-$) contributions. 
}
\label{fig:distributions}
\end{figure}

The probability density function (PDF) $p(J)$ of the instantaneous heat flux was obtained 
numerically and is presented in Fig.~\ref{fig:distributions}a, 
for different values of $\alpha$, 
chosen to illustrate cases yielding positive ($\alpha=1$), null ($\alpha=2$) and negative ($\alpha=3$) fluxes.
In the inset, we show that the ratio $p(J)/p(-J)$, which measures asymmetry, obeys a reversibility relation~\cite{crooks}, decaying exponentially in the limit of large fluxes.
Notice that small (large) negative fluxes are less (more) likely than  positive ones.

In Fig.~\ref{fig:distributions}b, we show how each value of $\alpha$ tunes 
the contributions of small and large scales of $J$ to the average flux 
$\langle J \rangle=\int_{-\infty}^\infty j\, p(j)\,\text{d}j$, that is decomposed as
\begin{equation}
\langle J \rangle  =\int_{0}^\infty j \, w(j)\,\text{d}j
=\underbrace{\int_{0}^{J_0} j\, w(j)\,\text{d}j}_{\mathcal{A}_+} + 
\underbrace{\int_{J_0}^\infty j\, w(j)\,\text{d}j}_{ \mathcal{A}_-} \, ,
\end{equation}
where $w(j) = j [p(j) - p(-j)]$ and $J_0$ is the point at which $w(j)$ becomes negative.
In all cases, while the short scales yield a positive bias  
$\mathcal{A}_+$ (light gray), the large scales yield a negative one $\mathcal{A}_-$ 
(dark gray). 
The positive contribution of the short scales indicates that the distributions are not 
symmetric near the origin, which goes almost unnoticed at nude eye. 
In the inset,  we can see how the relative contribution of short and large scales changes with $\alpha$, 
being responsible for flux inversion.
While the positive contribution (light gray bars) 
grows slowly with  $\alpha$  around $\alpha=2$, 
the negative contribution (dark gray bars) noticeably  increases.
This shows that the occurrence of large rare events is the feature 
being modulated by $\alpha$ that more strongly regulates flux direction.
Note that, even in the absence of net flux (for $\Delta T=0$ and $\alpha=2$), 
there is a non-equilibrium signature in the asymmetry of the PDF, 
as if there were a preferential positive direction given by the more 
likely small $J$ events, but which is ultimately compensated by strong rare negative ones.

\section{Heuristic considerations}
\label{sec:heuristic}
Our numerical results showed that, 
depending on the nonlinearity of the conducting medium, 
the role of Poisson higher-order  cumulants 
can change in a critical way, 
with inversion of its contribution to  heat flux direction. 
For FPUT chains, it has been previously shown through approximate methods 
that the higher-order cumulants generate 
a negative correction to heat flux~\cite{kanazawa2013,candido17}. 
Despite the fact that these calculations are performed under strong approximations, 
an educated guess of our results about flux inversion can be extracted as follows.

From the perspective of non-Gaussian stochastic energetics~\cite{kanazawa2012}, 
through a perturbation approach at the over-damped limit (neglecting inertial effects), 
with small $\Delta T$ and weak nonlinearity, it has been shown that the heat flux is given by  

\begin{equation} \label{fourier-general}
 J =  \sum_{n\ge 2} J_n  = - \sum_{n\ge 2} \kappa_n \Delta K_n \,,
\end{equation}
with $\Delta K_n =  K_n^{(R)} -K_n^{(L)}$,
and
\begin{equation} \label{kappa}
    \kappa_n = \frac{1}{2n!} \left \langle V^{(n)}(z) \right\rangle_{\text{eq}} \equiv		
		\frac{1}{2n!Z} \int_{-\infty}^\infty    V^{(n)}  e^{-V(z)/T} \text{d}z 
	\,,
\end{equation}
where   $V^{(n)}$ is the  $n$-th derivative of the potential and $Z$   the partition function. 
 Eq.~(\ref{fourier-general}) represents a generalization of Eq.~(\ref{fourier}).

In the harmonic case, the non-Gaussian contributions are turned off, 
since the series becomes truncated at $n=2$  due to null higher-order derivatives.
For arbitrary $\alpha$, the potential $V_\alpha(x)$ in Eq.~(\ref{eq:potential}) 
is in general non-analytic at $x=0$. 
However,  it can be cast in the regularized form 
\begin{equation}
\label{regular}
V_\alpha(x;x_0) = k [(x^2 + x_0^2)^{\alpha/2} - x_0^\alpha]/\alpha \,,
\end{equation}
with $x_0>0$.  
Substitution of $V_\alpha$, even in this regularized form, 
into Eq.~(\ref{kappa}) leads to nonconvergent series of Borel type. 
We bypass this issue, by directly checking the behavior of the coefficients 
$\kappa_n$ at the vicinity of the harmonic case, 
with the purpose of obtaining an indication of flux inversion. 
Then, we expand the potential around $\alpha=2 + \varepsilon$ (with $|\varepsilon| \ll 1$), 
namely,
$
\label{expansion-alfa}
V_\alpha(x;x_0)\approx  k x^2/\alpha + k[\varepsilon/\alpha]  f(x) + \mathcal{O}(\varepsilon^2)\,,
$
where $f(x)$ is independent of $\alpha$. 
At first order, the coefficients in Eq.~(\ref{kappa}) become 
$\kappa_n \propto \varepsilon$, 
flipping sign around $\alpha=2$ ($\varepsilon=0$). 
The change of sign  holds when summing up  
Eq.~(\ref{fourier-general}), producing the flux inversion phenomenon, 
with linear dependence  $J \propto (2-\alpha)$, 
as observed in Fig.~\ref{fig:gpflux}.
Importantly, note that the non-monotonic dependence on temperature 
reported on Fig.~\ref{fig:temp} is not captured by the over-damped 
(and weak nonlinear) approximation made to obtain Eq.~(\ref{kappa}), 
that after some calculations yields $|J| \sim T$ (for $\alpha=4$).

Through another perturbation approach, for an FPUT-chain interacting potential  
$V(x)= \frac{1}{2} k_1x^2 + \frac{1}{4}k_3 x^4$ (with $k_1 \gg k_3>0$), 
it has been shown~\cite{candido17} that the excess current $\delta J$ 
generated by the  higher-order  cumulants from the Poissonian 
bath $\mathcal{P}$, at first order in $k_3$, is 
\begin{equation} \label{k3}
\delta J  = \langle J \rangle_{\text{$\mathcal{G}$-$\mathcal{P}$}} - 
\langle J \rangle_{\text{$\mathcal{G}$-$\mathcal{G}$}} =  -{\mathcal C} k_3 \,,
\end{equation}
where the proportionality factor ${\mathcal C}>0$ 
depends on $k_1$, $\gamma$, $\lambda$ and $\bar{\Phi}$. 
That is, the Poisson character of the bath, interplaying with the quartic anharmonicity, 
gives a negative contribution to the net current, as if it were at an effective temperature which is 
higher than $T_R$. 
In our case, the regularized potential in Eq.~(\ref{regular}) admits a Taylor expansion near the origin
$ \label{taylor}
  V_\alpha(x;x_0)        
	= k \frac{x_0^\alpha}{2} \sum_{n=1}^\infty    \frac{\Gamma(\alpha/2)}{n!\Gamma(\alpha/2-n+1)}  
				\left(  \frac{x}{x_0} \right)^{2n} =      
 k\left[ \frac{1}{2} x_0^{\alpha-2} x^2 + \frac{1}{8} (\alpha-2) x_0^{\alpha-4} x^4 + \ldots \right],
$
which converges for $|x|<x_0$. 
When $x_0\to 0$ (or $x$ is large enough), the regularized potential recovers Eq.~(\ref{eq:potential}). 
Consistently, we observe that the numerical results are not significantly affected 
by the introduction of small $x_0$ (see Fig.~\ref{fig:gpflux}). 
However, we also observe that, when $x_0$  increases, smoothing the potential at the origin, 
the contributions of higher-order cumulants are reduced (see Fig.~\ref{fig:gpflux}).
This expansion allows to identify the effective coefficient $k_3 \simeq k(\alpha-2)x_0^{\alpha-4}/2$, 
that according to Eq.~(\ref{k3}) indicates a change of sign around $\alpha=2$.
Although the derivation of Eq.~(\ref{k3}) implicitly assumes that $k_3\ge0$ to have 
a confining potential, the mathematical derivation in the vicinity of $k_3=0$ 
is expected to hold independently of the sign of $k_3$, while higher-order terms 
would be responsible for the confinement.

In sum, from approximations developed for FPUT potentials 
in the vicinity of the harmonic case, we extracted information 
that suggests that the contribution of the higher-order cumulants 
to the flux, giving further support to the results from numerical simulations. 

\section{Final remarks}
\label{sec:final}
Previous studies of heat conduction that investigated the interplay between 
the cumulants of the baths 
and the nonlinearity of the propagation medium 
focused on the FPUT model~\cite{kanazawa2013,morgado16}. 
It was observed that the  higher-order cumulants, beyond the second order one,  
generate a (unidirectional) negative contribution to the heat flux,  
that is,  from the Poisson to Gaussian bath.  
Because the contribution of the quartic anharmonicity is one-way, 
only by exploiting a positive bias of the standard temperature  difference, 
it would be possible to produce current inversion.

Our proposal of a general power-law form for the potential, 
with arbitrary values of $\alpha$ beyond the quadratic and quartic cases, 
allows to unveil that,  
the non-harmonic correction to heat flow can be either negative or positive 
(for super-harmonic and sub-harmonic potentials,  respectively), 
 as pictorially represented in Fig.~\ref{fig:chain}. 
This effect allows a bidirectional control over heat transfer, 
keeping baths properties unchanged.
By deepening in heat flux statistics, we highlight the leading role of rare events.
Furthermore, we provide heuristic considerations that support our findings.

It is worth to recall that low-dimensional momentum (and stretch) conserving systems, subject to Gaussian (thermal) baths,  yield anomalous heat transport~\cite{lepri2003,dasdhar14}, 
particularly, presenting deviations from the scaling predicted by Fourier's law, 
$J \sim 1/N$~\cite{Das2014}.
This extends to our case where transport is solely led by non-Gaussian fluctuations ($\Delta T = 0$), but a rigorous characterization of size dependence  (not shown)
is computationally challenging and can be inconclusive even for extremely large chains, 
as discussed previously for the FPUT model under thermal baths~\cite{Das2014}. Despite that, the main result that we present, regarding the direction of flux for three-particle chains, persists for $N\gg 1$.
We also checked that the introduction of an on-site potential of the quartic form~\cite{phi4}, does not affect this main novel feature that we report.

 At last, it is interesting to notice that a device able to switch between interaction potentials of the sub- and super-harmonic types, or governed by a deformable potential 
able to undergo  a change  around $\alpha=2$, can control flux direction.
Beyond the heat conduction problem, this nonlinear control might have implications, for instance, in the performance and functioning of 
thermally driven systems, like nanomachines and Brownian motors~\cite{nlmachine,defaveri17,REIMANN,martinez2015}. 
 
\cite{eshuis2010,gnoli2013,kanazawa2015,sano2016}


\textit{Acknowledgments.---} We are grateful to Welles A. M. Morgado for initial discussions. This work is partially funded by the Spanish Research Agency, through grants ESOTECOS FIS2015-63628-C2-1-R (AEI/FEDER, EU)  and MDM-2017-0711 from the Maria de Maeztu Program for units of Excellence in R\&D (EHC); and by the Brazilian Research Agencies Coordena\c c\~ao de Aperfei\c coamento de Pessoal de N\'ivel Superior (CAPES), Conselho Nacional de Desenvolvimento Cient\'ifico e Tecnol\'ogico (CNPq), Funda\c c\~ao de Amparo \`a Pesquisa do Estado do Rio de Janeiro (FAPERJ).  
 
\bibliographystyle{apsrev4-1}
%

%

\end{document}